# Front propagation into unstable and metastable states in Smectic C* liquid crystals: linear and nonlinear marginal stability analysis.

Wim van Saarloos[1], Martin van Hecke[1] and Robert Holyst[2]

[1]*Institute Lorentz, Leiden University, P.O. Box 9506, 2300 RA Leiden, The Netherlands*
[2]*Institute of Physical Chemistry of the Polish Academy of Sciences, Dept. III, Kasprzaka 44/52, 01224 Warsaw, Poland*

We discuss the front propagation in ferroelectric chiral smectics (SmC*) subjected to electric and magnetic fields applied parallel to smectic layers. The reversal of the electric field induces the motion of domain walls or fronts that propagate into either an unstable or metastable state. In both regimes, the front velocity is calculated exactly. Depending on the field, the speed of a front propagating into the unstable state is given either by the so-called linear marginal stability velocity or by the nonlinear marginal stability expression. The cross-over between these two regimes can be tuned by a magnetic field. The influence of initial conditions on the velocity selection problem can also be studied in such experiments. SmC* therefore offers a unique opportunity to study different aspects of front propagation in an experimental system.

61.30G, 03.40K, 75.60G

In the ferro-electric smectic liquid crystal (SmC*), the polarization vector, perpendicular to the director and parallel to smectic layers, forms a helicoidal structure with the characteristic pitch (wavelength) of the order of micrometers. The director is tilted with respect to layers and precesses together with the local polarization about the axis perpendicular to the layers [1]. These systems are not only interesting from the scientific but also from the practical point of view [2], since they can be used as fast electro-optical switches. As noted in 1983 by Cladis *et al.* [3], for sufficiently large electric fields a description in terms of domains where the polarization is parallel to the field, separated by the domain walls, becomes appropriate. Inside the wall the director makes the full $2\pi$ twist and thus is in the unfavorable configuration relative to the electric field — see Fig. 1(a). The size of a domain is proportional to the pitch. When the field is reversed each wall splits into two domain walls that propagate in opposite directions into the domains where the director is pointing in the unfavourable direction — see Fig. 1(b). The reversal of the polarization is thus mediated by the propagation of domain walls or fronts, so that the switching time is proportional to the domain size and inversely proportional to the speed of the wall.

Inspired by the similarity in appearance of the dynamical equation for domain wall motion with the sine-Gordon equation, Cladis *et al.* [3] drew the analogy of this switching behaviour with the motion of solitons. Soon thereafter, however, Maclennan *et al.* [4] pointed out that since viscous effects are much larger than inertial effects in liquid crystals, the wall motion should actually be thought of as an example of front propagation into an unstable state. They then applied some results from the theory of front propagation into an unstable state [5] to the case in which the dielectric term can be neglected. In addition, they studied numerically how large the fields have to be for the domain wall picture to become applicable, and investigated the influence of dielectric [6] and backflow [7] effects.

Since the work of Maclennan *et al.* [4], the theory of front propagation into unstable states has been further developed [8–16]. In general, we know that there can exist two different types of regimes, that are often referred to as linear and nonlinear marginal stability [10,11]. In the linear marginal stability regime, the front speed, $v^*$, can be calculated explicitly from the dispersion relation of the unstable modes describing the dynamics of linear perturbations around the unstable state. Nonlinear marginal stability, on the other hand, refers to a regime in which the front speed, $v^\dagger$, is larger than $v^*$, and in which it depends on the fully nonlinear behavior of the equation. As a result, explicit calculations showing the presence of the nonlinear marginal stability regime are only available for a few equations [9,11–16].

It is the purpose of this paper to show that the above realization of domain wall or front propagation in SmC* liquid crystals provides an extremely interesting physical example of front propagation into unstable or metastable states:

*(i)* We show that an exact solution of the equation describing twist dynamics in SmC*, found by Cladis [17] and independently by others [18], is exactly the *nonlinear front solution* [12] that determines the velocity $v^\dagger$ in some parameter ranges.

*(ii)* So far nonlinear marginal stability has been established only for equations with polynomial nonlinearities [9,11–13,15]. Our analysis is done explicitly for an equation with nonpolynomial nonlinearities. Moreover, our results are a nice illustration of the observation [11] that nonlinear marginal stability often occurs near points where a cross-over to front propagation into a metastable state occurs.

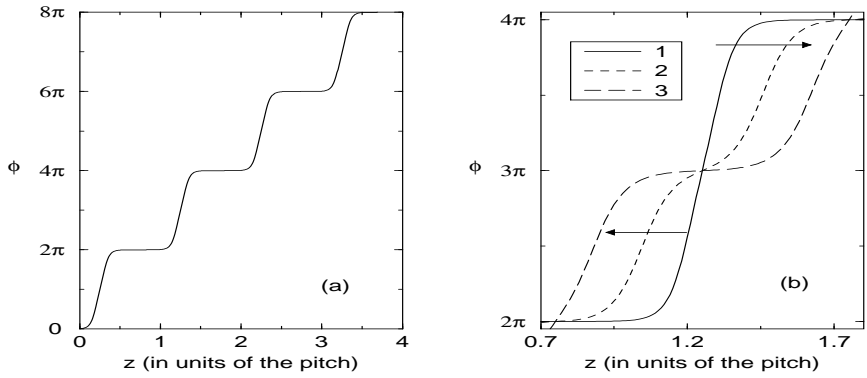

FIG. 1. (a) Stable stationary state of Eq. (4) for positive $E$, showing domains where $\phi = 0, \pm 2\pi, \ldots$ seperated by domain walls where $\phi$ makes a rapid twist of $2\pi$. (b) After $E$ is reversed, the $\phi = 0, \pm 2\pi, \ldots$ domains are invaded by $\phi = \pm \pi, \ldots$ states, as indicated by the arrows. The curves marked 1,2 and 3 show the initial state at $t = 0$ and two subsequent states after the reversal of $E$.

*(iii)* To our knowledge, front propagation in SmC* liquid crystals is the first physically realistic system where the cross-over from linear to nonlinear marginal stability appears accessible in experiments of the type of those performed by Cladis *et al.* [3].

*(iv)* If detailed experiments of the type indicated under *(iii)* are feasible, these experiments will also be the first ones in which the possibility arises to obtain front speeds larger than $v^*$ by preparing special initial conditions [5,9–11].

Let us consider a ferro-electric, chiral smectic system subjected to an electric field, $E$, and a magnetic field, $H$, parallel to the smectic layers and perpendicular to each other (as discussed later, the case with $H$ parallel to $E$ can be accounted for by a change in sign in $\chi_a$). The electric and magnetic energy density of the system is then given by [19,20]

$$F(\phi) = -\left(\frac{\epsilon_a E^2}{8\pi} + \frac{1}{2}\chi_a H^2\right)\cos^2\phi - PE\cos\phi , \qquad (1)$$

where $P$ is the polarization, and $\phi$ is the azimuthal angle between the electric field and polarization. For convenience, terms independent of $\phi$ have been omitted. The anisotropic part of the polarization is given by the three principal values of the polarization tensor and the tilt angle $\theta$ of the molecules in the smectic layers, i.e. $\epsilon_a = \epsilon_{pp} - \epsilon_\parallel \sin^2\theta - \epsilon_\perp \cos^2\theta$ [19,20]. For the diamagnetic anisotropy one similarly has $\chi_a = -\chi_{pp} + \chi_\parallel \sin^2\theta + \chi_\perp \cos^2\theta$. For simplicity we will concentrate on the case that both $\epsilon_a$ and $\chi_a$ are positive and discuss the major differences with the other cases briefly at the end. The sign of $\epsilon_a$ depends on the chemical structure of the constituent molecules [19]. If we take $E$ positive and $H$ fixed, then for $E_1 < E < E_2$, $F$ has a maximum at $\phi = \pm \pi, \pm 3\pi, \cdots$ and minima at $\phi = 0, \pm 2\pi, \cdots$, as shown in Fig. 2a. Here the cross-over fields are given by

$$E_1 = \frac{1}{2}E_c\left(1 - \sqrt{1 - H^2/H_m^2}\right) , \qquad E_2 = \frac{1}{2}E_c\left(1 + \sqrt{1 - H^2/H_m^2}\right) , \qquad (2)$$

where $E_c = 4\pi P/\epsilon_a$ and $H_m^2 = \pi P^2/\epsilon_a \chi_a$. For fields outside the above range, i.e. for $0 < E < E_1$ or $E > E_2$, there are additional local minima in $F$, as sketched in Fig 2b. Note that the free energy density $F$ is invariant under a reversal of the electrical field and change of $\phi$ by $\pi$. As a results, when the field direction is reversed ($E \to -E$) in the case of Fig. 2a, the global minima at $\phi = 0, \pm 2\pi, ..$ become maxima, as shown in Fig. 2c. For the case of Fig. 2b, however, the absolute minima at $\phi = 0, \pm 2\pi, ..$ change into local minima and the local minima at $\phi = \pm \pi, \pm 3\pi, ..$ change into global minima of $F$ under field reversal — see Fig. 2d. Suppose we start with a positive field $E$. As mentioned before, the fully relaxed SmC* will have large domains where $\phi \approx 0, 2\pi$, etc, separated by domain walls where $\phi$ changes rapidly by $2\pi$. If we now reverse the field, $E \to -E$, the states in these domains become unstable for $E_1 < E < E_2$ (the '*unstable field range*') and metastable for $E$ outside this range (the '*metastable field range*'). The stable domain walls that existed before the field reversal then become unstable and split into two fronts that propagate into these domains. Depending on the field strength, we thus have either a problem of front propagation into an unstable state or into a metastable state. The equation which governs the dynamics of the twist angle $\phi$ is [3,4,19]

$$\eta \frac{\partial \phi}{\partial t} = K \frac{\partial^2 \phi}{\partial z^2} - \frac{\partial F}{\partial \phi} \; , \qquad (3)$$

where $\eta$ is the rotational viscosity, and $K$ is the elastic constant. The $z$-axis is taken normal to the smectic layers. Since the propagation starts with a reversal of the field we consider a case in which a field $E$ is switched to $-E$. From here on we will measure time in units of $\eta/PE_c$, length in units of $\sqrt{K/PE_c}$, the electric field in units of $E_c$ and the magnetic field in units of $H_m$. In these rescaled units, the above dynamical equation then becomes with (1), after the field has been switched to the negative value $-E$

$$\frac{\partial \phi}{\partial t} = \frac{\partial^2 \phi}{\partial z^2} + E \sin \phi - \frac{1}{2}\left(E^2 + \frac{1}{4}H^2\right) \sin 2\phi \; . \qquad (4)$$

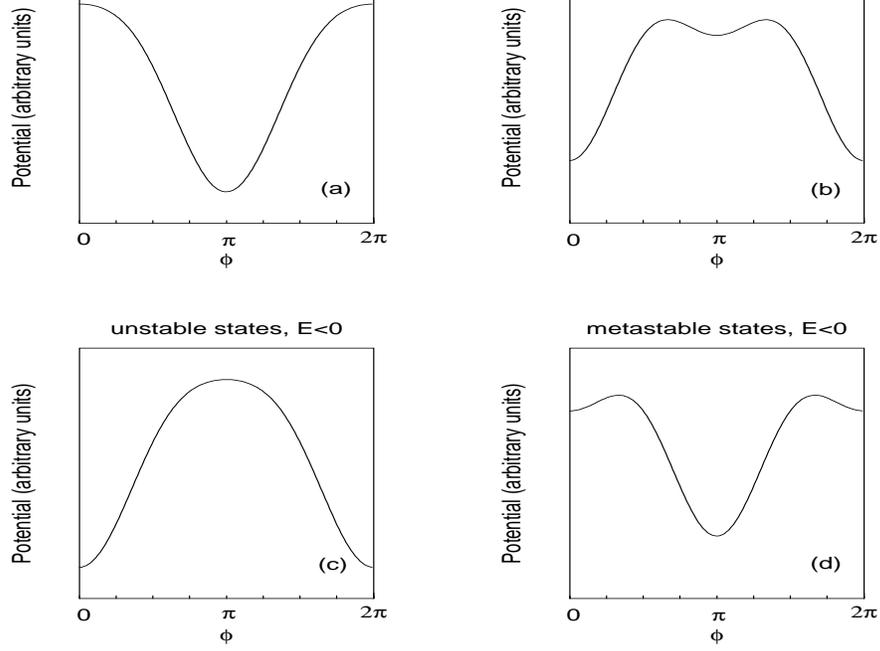

FIG. 2. The electrical free energy density (1) as a function of $\phi$. (a) and (c) show $F$ in the *unstable field range* for $E > 0$ and $E < 0$, respectively, and (b) and (d) in the *metastable field range*. $F(\phi)$ is a periodic function of $\phi$, but only the range $[0, 2\pi]$ is shown.

This equation is an example of a reaction-diffusion equation $\phi_t = \phi_{zz} + f(\phi)$ that has been studied extensively in the context of front propagation [5,9–11,13–16]. Surprisingly, the physical relevant case (4) can be solved exactly. Our main results concern explicit expressions for the asymptotic velocity of the single front propagating into the state $\phi = 0$ and creating a domain where $\phi = \pi$ (all modulo $2\pi$, of course). We will first derive these results, and then discuss the implications for the switching in SmC$^*$.

*Metastable field regime.* For equations of type (4), the front solution quickly approaches a uniformly traveling wave solution of the form $\hat{\phi}(z - vt) = \hat{\phi}(\xi)$; our goal is to determine $v$ for a front moving to the right into the state $\phi = 0$ at $\xi \to \infty$. When substituted into (4), the Ansatz $\phi = \hat{\phi}(\xi)$ leads to a single second order ordinary differential equation for $\hat{\phi}$. As this is equivalent to a set of two first order ordinary differential equations for $\hat{\phi}$ and $\hat{\phi}'$, we can think of this set of equations as describing a flow in a two-dimensional phase space. The metastable state at $(\hat{\phi} = 0, \hat{\phi}' = 0)$ into which the front propagates and the stable state $(\hat{\phi} = \pi, \hat{\phi}' = 0)$ created by the front correspond to fixed points of these equations. Linearizing around the metastable fixed point by substituting $\hat{\phi} \sim e^{-\lambda \xi}$, one finds

$$\lambda_{\pm} = \frac{1}{2}\left(v \pm \sqrt{v^2 - 4(E - E^2 - H^2/4)}\right) \; , \qquad (5)$$

In the metastable field regime, $\lambda_+$ is positive while $\lambda_-$ is negative. Thus, the perturbation $\exp(-\lambda_-\xi)$ diverges away from the fixed point $(\hat{\phi}=0, \hat{\phi}'=0)$, and so the required front solution must approach this fixed point for $\xi \to \infty$ along the single stable eigendirection $e^{-\lambda_+\xi}$. A straightforward analysis near the other fixed point shows that there is also only one appropriate eigendirection for the front solution there. These results together imply that the required front or domain wall solution corresponds to a single unique trajectory in phase space connecting the two fixed points (a so-called 'heteroclinic trajectory'), which only exists at a particular, unique value of the velocity.[1] As noted in [11,12,15], these special solutions can often be found by making the Ansatz that they are actually solutions of a first order differential equation $\hat{\phi}' = h(\hat{\phi})$, with $h$ a suitable function of $\hat{\phi}$. Here, the simple choice $h(\hat{\phi}) = -\lambda_+ \sin\phi$ is found to give the solution [17,18]

$$\hat{\phi}(\xi) = 2\arctan(\exp(-\lambda\xi)) \;, \text{ where } \quad v = \sqrt{\frac{E^2}{E^2 + H^2/4}}\;, \quad \lambda = \sqrt{E^2 + H^2/4}\;. \tag{6}$$

This result solves the asymptotic behavior of fronts in the *metastable field regime*.

*Unstable field regime.* In the unstable field regime, there are two positive real roots $\lambda_\pm$ according to (5) over some range of velocities. As is well known [5,9–12], the above phase space type arguments then imply that there is then a continuous set of solutions of the form $\hat{\phi}(x-vt)$, with $v$ in some range. The minimum value of the velocity range for which the roots $\lambda_\pm$ are real is according to (5)

$$v^* = 2\sqrt{E - E^2 - H^2/4}\;, \tag{7}$$

At this point the roots $\lambda_\pm$ coincide and are equal to

$$\lambda^* = \sqrt{E - E^2 - H^2/4} \tag{8}$$

Note that for an arbitrary velocity $v > v^*$, one expects the asymptotic decay of the front profile to zero to be governed by the smallest root $\lambda_-$, i.e. to have $\hat{\phi} \sim \exp(-\lambda_-\xi)$ as $\xi \to \infty$.

Since there is a whole set of steady state solutions, additional dynamical arguments are needed in the unstable field regime to determine the selected front speed. For equations of the simple type (4), the results from the theory [5,9–12] can be summarized as follows: $v^*$ (the so-called 'linear marginal stability' velocity) is the asymptotic front speed (for sufficiently localized initial conditions — see below) *unless* there exists a particular *nonlinear front solution* [11,12] with the property that it is faster and that its asymptotic decay rate $\exp(-\lambda^\dagger \xi)$ is not governed by the smallest root, but instead by $\lambda_+$:

$$v^\dagger > v^* \quad and \quad \lambda^\dagger = \lambda_+(v^\dagger) > \lambda^* \tag{9}$$

If such a front solution, which in technical terms is called a strong heteroclinic orbit [15], exists, then $v^\dagger$ is the selected front speed.

Physically, one expects that when upon varying a parameter the state into which a front propagates changes from metastable into unstable, the selected front speed will not change abruptly. This expectation is nicely embodied in the structural stability hypothesis underlying the recent approach by Paquette and Oono [13]. As noted in [11,12], in practice this means that the unique front solution found in the metastable regime, and whose asymptotic decay is governed by $\lambda_+$, becomes precisely the nonlinear front solution that satisfies (9) over some range of parameters in the metastable regime. In the present case, this expectation is borne out again: it is straightforward to verify that the solution given by (6) is indeed the nonlinear front solution satisfying (9) for $E < E_1^\dagger$ and $E > E_2^\dagger$, with

$$E_1^\dagger = \frac{1}{4}\left(1 - \sqrt{1-4H^2}\right)\;, \qquad E_2^\dagger = \frac{1}{4}\left(1 + \sqrt{1-4H^2}\right)\;. \tag{10}$$

---

[1] Strictly speaking, there is a discrete set of front solutions. Only the one with the largest velocity is stable, and this is the one we determine.

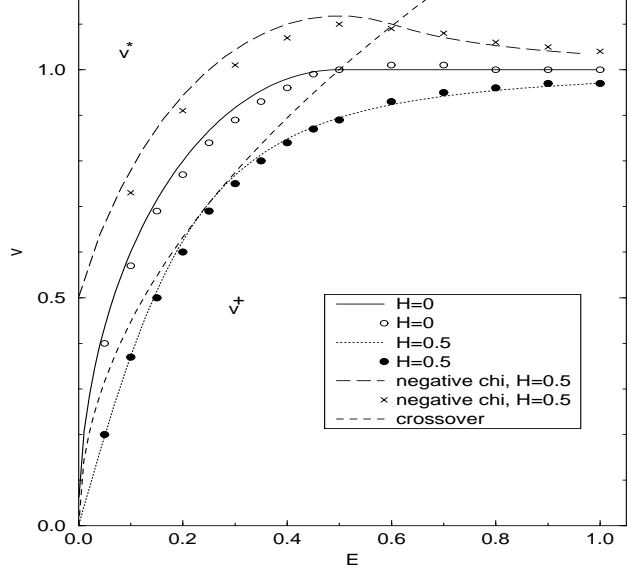

FIG. 3. Results for the predicted front velocity for three values of the magnetic field: $H=0$, $H=0.5$ with positive $\chi_a$ (in this case, $v = v^\dagger$ for all field values $E$) and $H=0.5$ with negative $\chi_a$, in which case $E_2^\dagger = (1+\sqrt{2})/4 \approx 0.6$. The symbols indicate the values obtained in numerical simulations in a finite system. The crossover between linear and nonlinear marginal stability is found from (7) or (8) and (11) to be $v_{cross} = \sqrt{2E}$. To the left of this curve, $v^*$ is selected, and $v^\dagger$ is found to the right of this curve.

At these cross-over fields, $v^\dagger = v^*$, so that the front velocity smoothly changes over to the linear marginal stability value $v^*$ at these field values. Note also that for $H=0.5$, $E_1^\dagger = E_2^\dagger$; consequently, for magnetic fields larger than this value the velocity is always given by $v^\dagger$ (6) for any value of the electric field. Thus our predictions for the front speeds are

$$\begin{bmatrix} E_1 < E < E_2 \\ \text{unstable regime} \\ \\ E < E_1 \text{ or } E > E_2 \\ \text{metastable regime} \end{bmatrix} \quad \begin{array}{l} E_1^\dagger < E < E_2^\dagger : \quad v = v^* = 2\sqrt{E - E^2 - H^2/4}\,, \\ \text{else}: \quad v = v^\dagger = \sqrt{E^2/(E^2 + H^2/4)}\,, \\ \\ v = v^\dagger\,, \end{array} \qquad (11)$$

where the cross-over field values $E_1$ and $E_2$ given by (2) are written in dimensionless units.

We have verified the above predictions for the front velocities by performing numerical simulations of (4) in a large but finite system to obtain the asymptotic selected front speeds. As shown in Fig. 3, our results are in good agreement with the analytical predictions: the fact that the measured velocities are slightly below the predicted ones is due to the slow convergence to the asymptotic value in the linear marginal stability regime [11].

So far, we have confined the analysis to the case that both $\chi_a$ and $\epsilon_a$ are positive. However, the sign of $\chi_a$ can be changed by rotating the magnetic field 90 degrees in the plane of the smectic layers, so when $\chi_a$ is positive in the setup we discussed before, it can be made negative by making the magnetic field directions parallel to the electric field direction [19,20], or vice versa. In our equations, a change of sign in $\chi_a$ can be incorporated by changing the sign in front of each $H^2$ term[2]. As shown also in Fig. 3, the front speed goes through a maximum in this case for $H \neq 0$. Note that in all cases, the dimensionless front velocity approaches the value 1 for large field strengths.

How easily can these predictions be tested experimentally? The experiments by Cladis et al. [3] have already demonstrated the feasibility of measuring the domain switching with crossed polarizers. For such experiments to be interpretable in terms of front dynamics, the width $W$ of the fronts has to be much less than the pitch $p_0$ [4]. As shown by (6) and (8), for dimensionless field strengths of order unity, the dimensional wall $W$ width is of order $\sqrt{K/PE_c} = \sqrt{K\epsilon_a/4\pi P^2}$. Hence $W$ can be made small by taking a material with small $K$ and/or large $P$. $K$ is typically of order $10^{-7}$ dynes or somewhat larger [4,19], $\epsilon_a$ can range from 0.1 to a value of order unity [19], while the polarization $P$ is typically of order 10 in Gaussian-cgs units but it can be as large as $10^3$ [19]. For the typical parameter values, we then have a wall width of the order of a $\mu m$ or less. Typically the pitch $p_0$ is of the order of $10 \mu m$. Hence by selecting appropriate materials with a large polarization, it appears to be possible to satisfy the condition $W \ll p_0$ experimentally.

---

[2]In this case $E_1$ and $E_1^\dagger$ are both negative and therefore irrelevant. Moreover, for field strengths $E < (-1 + \sqrt{1 + H^2/4})/2$, the minimum energy configuration is obtained for an angle $\phi$ between 0 and $\pi$. As discussed further by [6], we then do not have a problem of front propagation, and switching can be much faster.

In a stable configuration, the system consists of domains of the favourable configuration, separated by $2\pi$ domain walls a distance $p_0$ apart. Upon reversing the field, these walls split into two fronts that move apart. For $W \ll p_0$, the switching time then approaches $p_0/2v$ where $v$ is the asymptotic front velocity. We have investigated numerically how accurate this estimate is as a function of the ratio $W/p_0$. For a dimensionless domain size of 100, the switch times coincide with the predictions from the asymptotic front speed within 3%, when $E$ is larger than 0.2; for smaller $E$ the wall size is relatively large and the discrepancy between asymptotic speed and switch times increases rapidly. In dimensionless units, the size of the domain walls for $E \gtrsim .2$ is of order 5.

For the typical values given above, the field scale $E_c = 4\pi P/\epsilon_a$ is of order 100 in Gaussian-cgs units, i.e. of order $3.10^4\ V/cm$. This values is right in the middle of the range of field values studied by Cladis et al.. Moreover, since the dimensionless velocity approaches 1 in the limit of large fields, we find with the parameter values recommended in [4,21] for DOBAMBC used in this experiment, a large field front velocity of $\eta^{-1}\sqrt{KPE_c} \approx 0.6\ cm/s$. With a pitch of 1.75 $\mu m$, this leads to a switching time of the order of 0.2 $ms$. This is comparable to the shortest switching time observed in the experiments of Cladis et al. [3], but there is no indication in their data that the switching time saturates in this range. Moreover, the condition $W \ll p_0$ is not very well satisfied in their experiments, and as noted by Maclennan et al. [4], when the dielectric terms are neglected completely, the switching is predicted to be faster than actually observed. Clearly, detailed experiments with appropriately selected materials will be needed to put our predictions to a stringent test.

Our analysis shows that in order to influence the front speed appreciable with a magnetic field, one needs fields of the order of $H_m = \sqrt{\pi P/\epsilon_a \chi_a}$. Using the typical value $10^{-7}$ for $\chi_a$ [19], $H_m$ is of the order of a Tesla. In an actual experiment, one also has to make sure that the fields do not exceed the critical values above which the helix may unwind [19,22], even though this can be a dynamically very slow process [22]. The typical field values that one needs in experiments on front motion turn out to be below this critical value. Unwanted effects from the motion of free ions can be prevented by using ac-field with a square envelope of frequency much lower than the inverse switching time [20].

In our analysis, we have also neglected backflow effects. If the viscosity is large, backflow effects can be neglected [7]. Zhou et al. [7] have also shown that for small fields backflow is not important even for small viscosity; for typical values of interest to us, we estimate following [7] that these effects are relatively unimportant.

We note one final remarkable point. In the theory of front propagation into unstable states, it is known [5,9–11] that the front speed can theoretically exceed $v^*$ in the linear marginal stability regime if the initial conditions are such that initially $\phi(x,t) = 0$ drops off as $e^{-\lambda_0 x}$ with $\lambda_0 < \lambda^*$. We do not know of any experimental system where one has been able to test this. If more detailed experiments similar to those of Cladis et al. can be done, they will in principle yield a way to test this by making the field strength before and after the field reversal different: the latter affects $\lambda^*$, the former $\lambda_0$.

In summary, we have calculated the front propagation in the SmC$^*$ phase subjected to electric and magnetic fields parallel to smectic layers. We showed that this system offers unique opportunities for observing the crossover between the linear and nonlinear marginal stability front propagation, and between front propagation into unstable states and into metastable states. Theoretically our results are of interest because they concern a rare case in which front propagation can be solved for nonpolynomial nonlinearities.

WvS is grateful P. E. Cladis for previous collaborations that have seeded this research [17]. The work by RH was supported in part by the Komitet Badań Naukowych (KBN) under grant No 2 P302 190 04 and 2 P303 020 07.